\documentclass[%
reprint,
superscriptaddress,
amsmath,amssymb,
aps,
prb,
floatfix,
]{revtex4-2}

\usepackage[caption=false]{subfig}
\usepackage{multirow}
\usepackage{graphicx}
\usepackage{dcolumn}
\usepackage{bm}
\usepackage{threeparttable}
\usepackage{booktabs}
\usepackage{color}
\usepackage{booktabs}
\usepackage{xspace}
\usepackage{chemformula}
\usepackage{xcolor}

\definecolor{myblue}{RGB}{0, 0, 0}

\newcommand{\CBO}{\mbox{\ch{CuB2O4}}\xspace}

\begin{document}

\title{
Magneto-toroidal nonreciprocity of second harmonic generation}

\author{J.~Mund}
\affiliation{Experimentelle Physik 2, Technische Universit{\"a}t Dortmund, D-44221 Dortmund, Germany}
\author{D.~R.~Yakovlev}
\affiliation{Experimentelle Physik 2, Technische Universit{\"a}t Dortmund, D-44221 Dortmund, Germany}
\affiliation{Ioffe Institute, Russian Academy of Sciences, 194021 St.-Petersburg, Russia}
\author{A.~N.~Poddubny}
\affiliation{Ioffe Institute, Russian Academy of Sciences, 194021 St.-Petersburg, Russia}
\author{R.~M.~Dubrovin}
\affiliation{Ioffe Institute, Russian Academy of Sciences, 194021 St.-Petersburg, Russia}
\author{M.~Bayer}
\affiliation{Experimentelle Physik 2, Technische Universit{\"a}t Dortmund, D-44221 Dortmund, Germany}
\affiliation{Ioffe Institute, Russian Academy of Sciences, 194021 St.-Petersburg, Russia}
\author{R.~V.~Pisarev}
\email{pisarev@mail.ioffe.ru}
\affiliation{Ioffe Institute, Russian Academy of Sciences, 194021 St.-Petersburg, Russia}

\date{\today}

\begin{abstract}
The Lorentz reciprocity principle is a fundamental concept that governs light propagation in any optically linear medium in zero magnetic field.
Here, we demonstrate experimentally a novel mechanism of reciprocity breaking in nonlinear optics driven by the toroidal moment.
Using high-resolution femtosecond spectroscopy at optical electronic resonances in the magnetoelectric antiferromagnet \CBO,
we show that by controlling the nonlinear interference of coherent sources of second harmonic generation originating from the toroidal spin order, applied magnetic field, and noncentrosymmetric crystal structure, we induce a huge nonreciprocity approaching 100\% for opposite magnetic fields.
The experimental results are corroborated by a convincing theoretical analysis based on the magnetic and crystal symmetry.
These findings open new degrees of freedom in the nonlinear physics of electronic and magnetic structures and pave the way for future nonreciprocal spin-optronic devices operating on the femtosecond time scale.
\end{abstract}

\maketitle

{A magnetic field induces optical nonreciprocity in any linear medium enabling one-way routing of light~\cite{caloz2018electromagnetic,potton2004reciprocity,zvezdin1997modern}, which
can be revealed by studying the Faraday rotation or magnetic circular dichroism, as sketched in Fig.~\ref{fig:fig1}a.
In nonlinear optics, the time-reversal symmetry is inherently broken and the principle of reciprocity is not valid even in the absence of a magnetic field~\cite{shen2003principles,boyd2008nonlinear,arima2019}. 
Nonlinear nonreciprocal effects become especially rich when complemented by the breaking of spatial inversion symmetry and nontrivial spin order~\cite{ogawa2004magnetization,spaldin2008toroidal,fiebig1994second,fiebig2002observation,fiebig2005second,spitzer2018routing}.
Despite rapid recent progress~\cite{tokura2018nonreciprocal,cheong2018broken,cheong2019sos}, this field has remained mostly uncharted so that novel, non-trivial effects from nonlinear and nonreciprocal resonant light-matter interactions due to the coupling of charge, orbital and spin degrees of freedom can be expected.

Here, we demonstrate widely tunable nonlinear reciprocity breaking due to interference of symmetry-different optical second harmonic generation (SHG) sources induced by an applied magnetic field $\bm B$ and a toroidal moment $\bm T$, as shown schematically in Fig.~\ref{fig:fig1}b.
The toroidal order, initially introduced in high-energy physics~\cite{dubovik1990toroid}, currently attracts tremendous attention across disciplines ranging from spin physics of multiferroics and magnetoelectrics~\cite{spaldin2008toroidal,tokura2018nonreciprocal,arima2019,lehmann2019poling,van2007observation,ederer2007towards,zimmermann2014ferroic,toledano2015primary,toledano2011spontaneous} to nanoscale optics~\cite{miroshnichenko2015nonradiating}.
However, the 
giant nonlinear magneto-toroidal nonreciprocity discovered here through the contribution 
\begin{equation}\label{eq:BT}
|E^{2\omega}(\bm B)|^2-|E^{2\omega}(-\bm B)|^2\propto  B_xT_x|E_z^{\omega}|^4
\end{equation}  
to the SHG intensity, has never been reported before to the best of our knowledge.
Our findings are based on high-sensitivity and high-resolution femtosecond spectroscopy of resonant optical SHG from the antiferromagnetic copper metaborate \CBO and corroborated by a rigorous symmetry analysis.
 
The choice of \CBO is motivated by its favorable combination of exceptionally narrow optical resonances and magnetic phase diagrams with different types of spin ordering.
This opens up yet unexplored opportunities for disclosing new mechanisms of nonlinear nonreciprocity.
We focus on the SHG at the electronic resonance near 1.405\,eV (Fig.~\ref{fig:fig1}c), taking place between the $3d^{9}$ states of the Cu$^{2+}$ ions at the $4b$ Wyckoff positions~\cite{martinez1971crystal}. 
While SHG at this transition was observed before~\cite{pisarev2004magnetic}, it has so far not been studied with respect to SHG nonreciprocity and the underlying mechanisms.
 
Here we focus on the nonreciprocity due to the coupling between the 
spins $S=1/2$ of the Cu$^{2+}$ ions that provide a rich variety of commensurate and incommensurate spin structures~\cite{boehm2003complex}.  The nonvanishing antiferromagnetic $\bm L=\bm S_{1}-\bm S_{2}$ and ferromagnetic $\bm M=\bm S_{1}+\bm S_{2}$ order parameters (Fig.~\ref{fig:fig1}b) characterize the commensurate phases. 
Remarkably, the order parameter $\bm L$ not only characterizes the commensurate phases   but carry also a nonzero toroidal vector $\bm T\propto L_x\hat{\bm y}+L_y\hat{\bm x}$. It is the interference of
the SHG activated by the toroidal moment $ \bm T $ and the SHG induced by the magnetic field $B$, that drives the strong magneto-toroidal SHG nonreciprocity according to
Eq.~\eqref{eq:BT}.
We achieve full control of sign and magnitude of this nonreciprocity by changing the direction and magnitude of the magnetic field. Additional nonreciprocity pathways
are offered by the spectrally-broad crystallographic SHG source due to the noncentrosymmetric crystal structure. Its controlled involvement provides SHG spectra containing pronounced Fano-resonances with strong spectral asymmetry as the hallmark of nonlinear interference.
Obviously, the interference of SHG sources, illustrated  in Fig.~\ref{fig:fig1}(c), is not limited to  \CBO, but can be used as new concept for magnetic routing of the nonlinear on- and off-resonant coherent emission in complex media.

\begin{figure*}
\centering
\includegraphics[width=1.75\columnwidth]{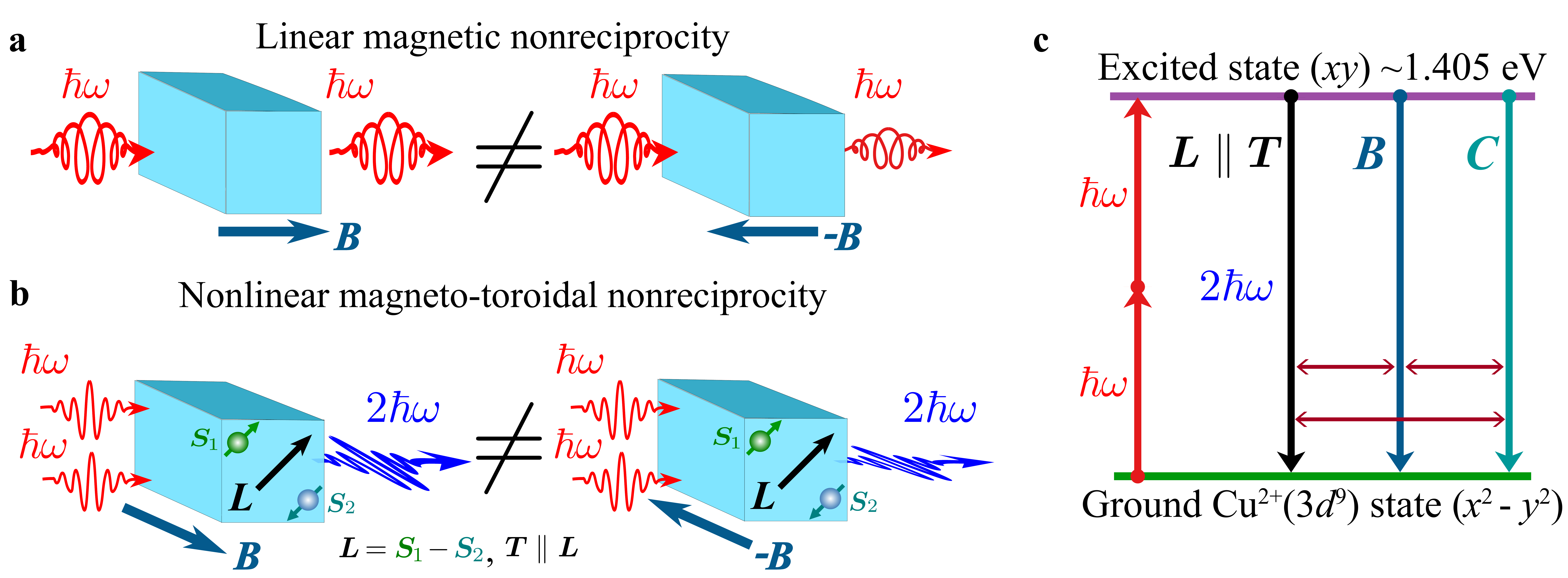}
\caption{\label{fig:fig1}
\textbf{Nonreciprocal second harmonic generation in \CBO.}
\textbf{a.} Nonreciprocity of light in a linear medium due to different absorption for propagation along or opposite to the applied magnetic field (magnetic circular dichroism).
\textbf{b.} Nonreciprocal second harmonic generation induced by the magnetic field $\bm B$ and the antiferromagnetic moment $\bm L = \bm S_1 - \bm S_2$ that is parallel to the toroidal moment $\bm T$.
\textbf{c.}~Electronic transition between the ground $(x^{2}-y^{2})$ state and lowest excited $(xy)$ state of Cu$^{2+}$ $4b$ ions near 1.405\,eV~\cite{pisarev2004magnetic}. 
These states are subject to Davydov splitting (not shown) due to the presence of two Cu$^{2+}$ ions at 4$b$ sites in the primitive unit cell~\cite{boldyrev2015antiferromagnetic}.
The SHG process converts two photons 
with frequency $\omega$ into one photon at frequency $2\omega$. 
The vertical lines in Fig.\,1c mark the crystallographic ($C$), magnetic-field-induced ($\bm B$), and toroidal ($\bm T\parallel\bm L$) SHG sources.
Interference pathways between these sources are indicated by the horizontal double-arrow lines.}  
\end{figure*}

\section*{Nonlinear magneto-toroidal nonreciprocity}

\begin{figure*}
\centering
\includegraphics[width=2\columnwidth]{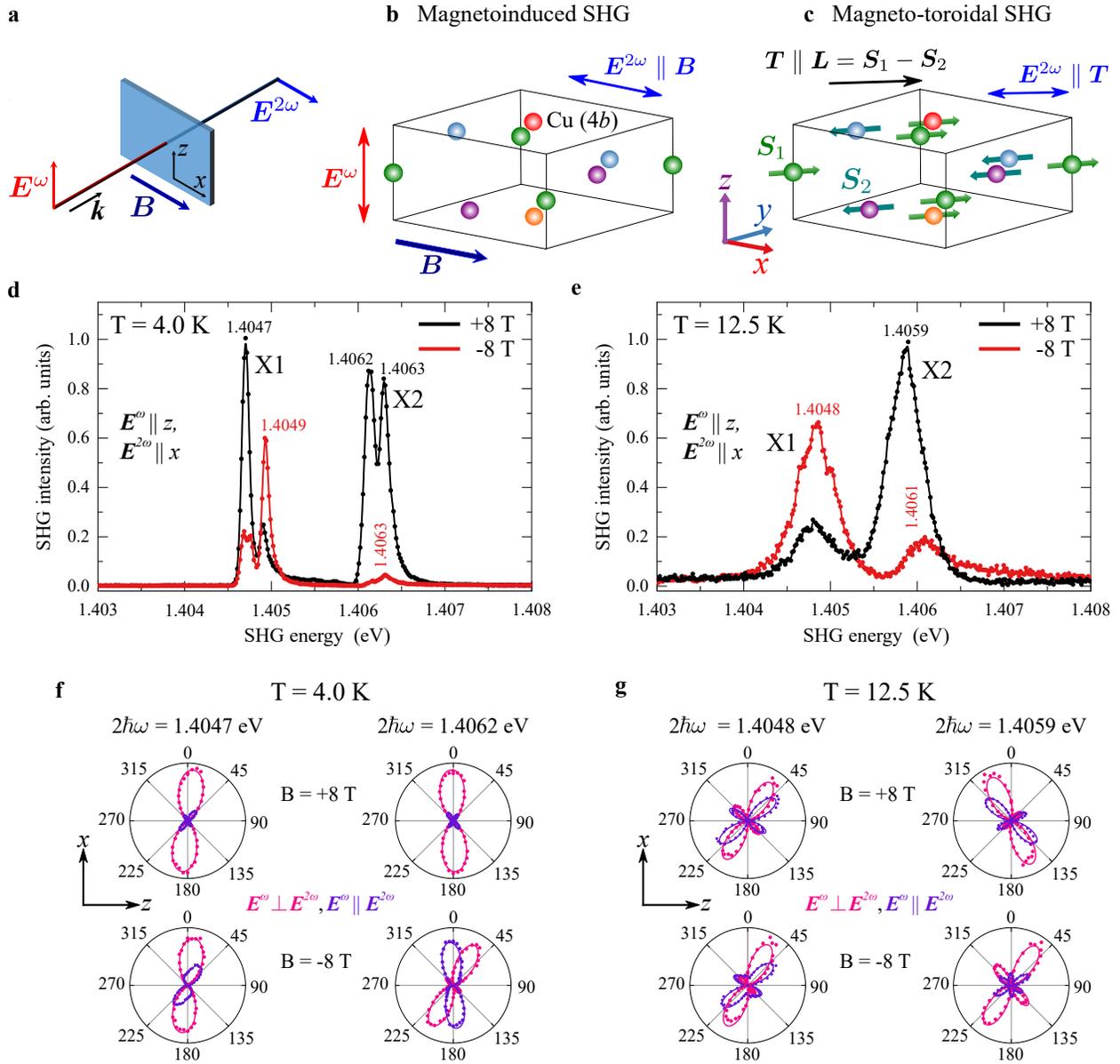}
\caption{\label{fig:fig2}
\textbf{Nonreciprocal SHG spectra of \CBO measured at normal incidence.}
\textbf{a.} Geometry of the experiment,  $\bm k \parallel y$ and $\bm B \parallel x$.
\textbf{b,c.} Polarization $E^{2\omega}$ of the SHG with respect to the crystal unit cell, induced by the magnetic field (\textbf{b}) and by the toroidal moment $\bm T\parallel \bm L$ (\textbf{c}).
Only Cu$^{2+}$ ions at the $4b$ positions and their spins are shown; ions of the same color have the same $z$ coordinate.
\textbf{d.} SHG spectra at $T=4.0$\,K for opposite fields which demonstrate strong spectral nonreciprocity reaching almost 100\% at some lines, e.g., at 1.4062\,eV line. 
\textbf{e.} SHG spectra at $T=12.5$\,K measured in the same geometry as in Fig.\,2d. Temperature increase from $T=4.0$ to 12.5\,K leads to line broadening, drop of the SHG intensity by about an order of magnitude (see Figs.\,3b and 3c) but the spectral nonreciprocity remains well pronounced. 
\textbf{f.} Rotational anisotropies at $T=4.0$\,K for opposite field directions in $\bm E^{\omega} \perp \bm E^{2\omega}$ (magenta symbols) and $\bm E^{\omega} \parallel \bm E^{2\omega}$ (blue symbols) for the two SHG lines at 1.4047 and 1.4062\,eV. These anisotropies are different for opposite fields demonstrating another side of nonreciprocity.
\textbf{g.} The same as in \textbf{f} but at $T=12.5$\,K for the two SHG peaks at 1.4048 and 1.4059\,eV.
These rotational anisotropy diagrams are noticeably different from those at 
$T=4.0$\,K shown in Fig.\,2f. 
We note that rotational diagrams for opposite fields at $T=4.0$ and 12.5\,K are similar for the first group of lines but very different for the the second group.
All rotational diagrams were fitted using equations based on the relevant SHG tensor components given in the Supplementary Section S3.  
}
\end{figure*}

We now present experimental results for the SHG nonreciprocity in two experimental geometries,
differing by the applied magnetic field orientation $\bm B$ and the resulting antiferromagnetic spin phase.


\subsection*{Voigt geometry $\bm B \parallel x$, $\bm k \parallel y$} 
Figure~\ref{fig:fig2} shows SHG spectra in the Voigt geometry when $\bm{B}\parallel{}x$ and the light wave vector $\bm{k}\parallel{}{y}$ (Fig.~\ref{fig:fig2}a).
The SHG spectra in Fig.~\ref{fig:fig2}d are recorded at $T=4.0$\,K in magnetic fields of $\pm$8\,T when the antiferromagnetic structure is in the commensurate phase 
(see phase diagram in Fig.~\ref{fig:fig3}a).
The spectra can be divided into two groups 
of sharp lines at 1.4047--1.4049 and 1.4062--1.4063\,eV with narrow full widths at half maximum (FWHM) below 100\,$\mu$eV.
We note that no line splitting could be observed in previous studies of \CBO when the SHG was excited by nanosecond pulses~\cite{pisarev2004magnetic}. 
The observed sharp SHG lines demonstrate the advantages of the femtosecond-pulse technique combined with high spectral resolution (see Methods). The two groups are separated by the Zeeman splitting of about 1.5\,meV. Each group demonstrates a small splitting which we assign to the
Davydov splitting~\cite{boldyrev2015antiferromagnetic}.
The detailed magnetic field dependence of the SHG spectra from 0 to $\pm$10\,T is discussed in the Supplementary Section S4.

The main result illustrated in Fig.~\ref{fig:fig2}d is the strong difference between the SHG spectra for oppositely oriented magnetic fields, compare black and red spectra.
This difference gives unambiguous evidence of strong SHG nonreciprocity, whose degree varies for the different lines and approaches 100\% at some photon energies, e.g., for the 1.4062\,eV line.
With increasing temperature up to $T=12.5$\,K (Fig.~\ref{fig:fig2}e) the SHG lines broaden, but the splitting into two groups and the nonreciprocity remain well pronounced,  .
 

The SHG nonreciprocity is further confirmed by the rotational anisotropies for the $\bm E^{\omega}\perp{}\bm E^{2\omega}$ (magenta symbols) and $\bm E^{\omega}\parallel{}\bm E^{2\omega}$ (blue symbols) polarization configurations at the 1.4047 and 1.4062\,eV lines ($T=4.0$\,K) in Fig.~\ref{fig:fig2}f and the 1.4048 and 1.4059\,eV lines ($T=12.5$\,K) in Fig.~\ref{fig:fig2}g.
Both series of measurements were performed with the magnetic field $B=\pm$8\,T, the observed differences can be explained by the difference in the field $\bm B\parallel{}x$ effect on the rearrangement of the antiferromagnetic spins in the $(xy)$ plane.
In zero field, the antiferromagnetic $\bm L$ vector is directed along the easy $[110]$ axis which is the diagonal between the $x$ and $y$ axes. But as the $B_x$ field increases, the $\bm L$ vector gradually rotates toward the $y$ axis perpendicular to the field direction. 
This process is controlled by sample temperature and magnetic-field
strength~\cite{petrova2018copper}.


\begin{figure*}
\centering
\includegraphics[width=2\columnwidth]{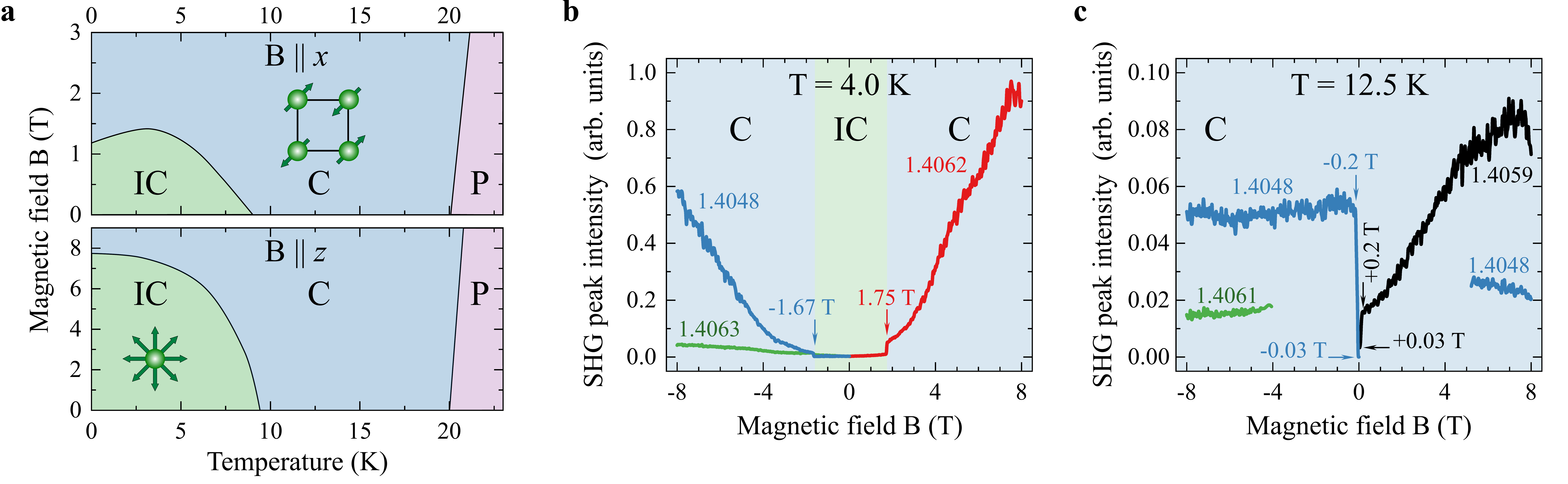}
\caption{\label{fig:fig3}
\textbf{Magnetic field dependences of SHG lines originating from the electronic transitions in the vicinity of 1.405\,eV in \CBO.} 
\textbf{a.} Schematic magnetic phase diagrams with the field $\bm B \parallel x$ and $\bm B \parallel z$.
The designations IC, C, and P refer to the incommensurate, commensurate, and paramagnetic phases.
More details related to the phase diagrams can be found in~\cite{boehm2003complex,fiebig2003magnetic,pisarev2004magnetic,martynov2007frustration,kousaka2007chiral,petrova2018copper,kawamata2019thermal}. 
\textbf{b}. At $T=4.0$\,K, no SHG signals are observed for the magnetic field in the range $-1.67<B<1.75$\,T when \CBO is in the incommensurate phase. 
Outside this field range, SHG lines appear and their field dependences are different for opposite field directions evidencing nonreciprocity. 
\textbf{c}. At $T=12.5$\,K, SHG signals appear already at magnetic fields as small as $\pm{0.03}$\,T when the antiferromagnetic domains disappear. Further increase of the field shows that each SHG line demonstrates a specific nonreciprocity as a function of the field.
}
\end{figure*}

The observed nonreciprocity is explained by Eq.~\eqref{eq:BT}, describing 
the interference of the two SHG sources due to the applied magnetic field,
$E_x^{2\omega}\propto ( E^{\omega}_{z})^2 B_x$  and the  toroidal moment $\bm T$,
$\bm E^{2\omega}\propto  ( E^{\omega}_{z})^2\bm T $, see Fig.~\ref{fig:fig2}(b,c). The  toroidal vector,
defined for a localized spin distribution as $\bm T=\frac{1}{2} \sum_j\bm r_j\times \bm S_j$, in  \CBO  is proportional to $L_y\hat{\bm x}+L_x\hat{\bm y}$. It is aligned with the $[110]$ axis, similar to  the antiferromagnetic moment $\bm L$ in the absence of a magnetic field, as shown in Fig.~\ref{fig:fig2}c. 
More formally, the observed SHG is due to the nonlinear magnetoelectric susceptibility $C^{L}_{xzzy}=C^{L}_{yzzx}$, and nonlinear magnetic-field-induced susceptibility  $C^{B}_{xzzx}=-C^{B}_{yzzy}$, 
while the crystallographic SHG vanishes in the considered geometry. 
Details of our symmetry analysis are given in the Supplementary Section S2. 

The magnetic phase diagram in Fig.~\ref{fig:fig3}a shows that orientation of the antiferromagnetic vector $\bm L$ and the toroidal moment $\bm T$ is
very sensitive to the direction and magnitude of the applied magnetic field.
According to Eq.~\eqref{eq:BT}, 
changes of the magnetic structure in the applied field should be reflected in the SHG field dependences. 
Figure~\ref{fig:fig3}b shows such field dependences of the SHG intensity for several lines at $T=4.0$\,K, where each line demonstrates its particular dependence.
When \CBO is in the incommensurate phase area marked by arrows between $-1.67 < B < 1.75$\,T  in Fig.~\ref{fig:fig3}b, no SHG signals are detected.
This observation confirms that rotation of the antiferromagnetic vector $\bm L$ in the $(xy)$-plane, while propagating along the $z$ axis~\cite{boehm2003complex}, results in such averaging of its SHG contribution when no signal is observed. 
Put in other words, the incommensurability destroys the nonreciprocity despite the applied magnetic field.
The SHG signal jumps from zero to finite values only after the phase boundary between the incommensurate and commensurate phases is crossed (see Fig.~\ref{fig:fig3}a).
The different values of the corresponding positive and negative critical fields are characteristic for the first order phase transition.
Further field increase leads to gradual rotation of the antiferromagnetic vector $\bm L$ in the $(xy)$ plane, resulting in strong SHG intensity changes as discussed above. 
Field dependences of the SHG lines at $T=12.5$\,K are shown in Fig.~\ref{fig:fig3}c.
They are drastically different from the corresponding dependences at $T=4.0$\,K (Fig.~\ref{fig:fig3}b). 
First of all, the SHG intensity decreases by about a factor of ten (compare the intensity scales in both Figures).
In contrast to Fig.~\ref{fig:fig3}b, the ``dead zone'' of zero SHG intensity and nonreciprocity is fully suppressed and SHG signals are observed in fields larger than $\pm{0.03}$\,T when the antiferromagnetic domains disappear.
The SHG field dependences are strongly nonreciprocal and each particular line demonstrates different behavior. 
Pronounced changes of the rotational anisotropies with temperature are observed when comparing Figs.~\ref{fig:fig2}f and~\ref{fig:fig2}g.

The entirety of spectral, magnetic field and rotational anisotropy results at $T=4.0$ and 12.5\,K in the chosen experimental geometry ($\bm k\parallel{}y$ and $\bm B\parallel{}x$) provides unambiguous evidences of SHG nonreciprocity, when \CBO is in the antiferromagnetic commensurate phase.

\begin{figure*}
\centering
\includegraphics[width=2\columnwidth]{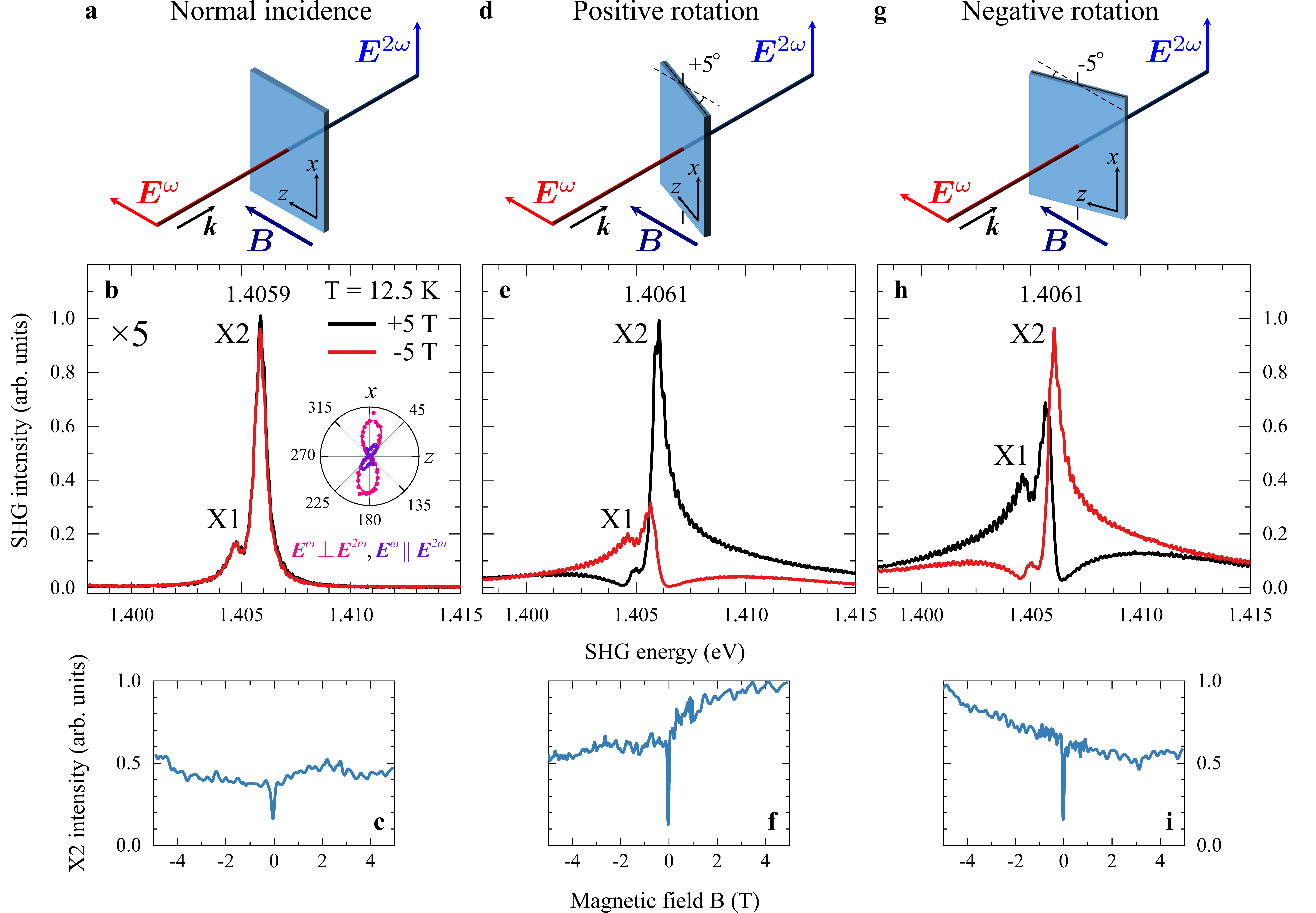}
\caption{\label{fig:fig4}
\textbf{SHG spectra of \CBO at normal incidence and for the sample rotated left and right  around the $x$ axis.} 
\textbf{a.} Reciprocal SHG spectrum measured at normal incidence for the fundamental light propagating along the $y$ axis, $\bm{k}\parallel{}y$, for the magnetic field $B=\pm{5}$\,T in the Voigt geometry, $\bm{B}\parallel{}z$.
Inset shows SHG rotational anisotropy of the $X2$ peak for $\bm{E}^{\omega}\perp{}\bm{E}^{2\omega}$ (red symbols) and $\bm{E}^{\omega}\parallel{}\bm{E}^{2\omega}$ (black symbols).
The rotational anisotropies for the $X1$ peak are similar to those of the $X2$ peak (not shown), in contrast to the $\bm{B}\parallel{}x$ geometry where the two groups showed different behavior (see Fig.~\ref{fig:fig2}).
Figure 4c shows the SHG intensity as a function of magnetic field, demonstrating saturation in a field as small as $\sim$50\,mT. A small signal in zero field is explained by incomplete compensation of opposite antiferromagnetic domains.
\textbf{b.} SHG spectra for the sample rotated around the $x$ axis by 5$^{\circ}$ angle. 
The resulting rotation enables crystallographic SHG contributions and increases the SHG signal by about a factor  about five in comparison to Fig.\,4a. 
The interference between resonant and non-resonant contributions results in nonreciprocal Fano-type SHG spectra differing for opposite magnetic fields.   
\textbf{c.} Same as \textbf{b.} but for the opposite rotation direction.
}
\end{figure*}

\subsection*{Voigt geometry $\bm{B}\parallel z$, $\bm{k}\parallel y$}
We now discuss SHG results for the same $(xz)$-plane sample at $T=12.5$\,K, also for $\bm k\parallel{}y$ in the Voigt geometry, but for the magnetic field $B=\pm{5}$\,T 
applied along the $z$ axis (Fig.~\ref{fig:fig4}a) where
\CBO is in the commensurate phase (Fig.~\ref{fig:fig3}a).
The strongest SHG signal (Fig.~\ref{fig:fig4}b) is observed for the polarizations of the incident light $\bm E^{\omega}\parallel{}z$ and the SHG light $\bm E^{2\omega}\parallel{}x$, similar to  Fig.~\ref{fig:fig2}e. The $X1$ and $X2$ peaks in Fig.~\ref{fig:fig4}b can be assigned to the two Davydov components~\cite{boldyrev2015antiferromagnetic} similar to Figs.~\ref{fig:fig2}d and~\ref{fig:fig2}e.
As noted above, the crystallographic contribution vanishes at normal incidence $\bm k\parallel{}y$ and therefore the SHG spectra in Fig.~\ref{fig:fig4}b definitely have to be assigned to the antiferromagnetic order. In contrast to the $\bm B\parallel{}x$ geometry no field dependence is observed above the saturation field (Fig.~\ref{fig:fig4}c). 
The role of the magnetic field in this case is just to reach saturation of the antiferromagnetic polarization.

In striking contrast to the $\bm B\parallel{}x$ results in Fig.~\ref{fig:fig2}, where the SHG nonreciprocity for opposite field directions reaches values as high as 100\%, the SHG spectra in Fig.~\ref{fig:fig4}b are exactly the same for opposite fields and no nonreciprocity is observed. Moreover, no difference is found in the rotational anisotropies for positive and negative fields of $B=\pm{5}$\,T. These results are unambiguously explained on the basis of the detailed symmetry analysis presented in the Supplementary Section S2.
In the given geometry at normal incidence, when $\bm k\parallel{}y$, $\bm B\parallel{}z$,
$\bm E^{\omega}\parallel{}z$, and $\bm E^{2\omega}\parallel{}x$, the SHG signal can be only induced due to the antiferromagnetic order because $P^{2\omega}_{x} = C^{L}_{xzzy} E^{\omega}_{z} E^{\omega}_{z} L_{y}$.
Thus, no interference and no nonreciprocity are allowed since the other contributions to the SHG are symmetry-forbidden.
The comparison between the $\bm B\parallel{}x$ and $\bm B\parallel{}z$ geometries
demonstrates that the antiferromagnetic $\bm L$ contribution to the SHG is observed in both cases.
However, it is nonreciprocal in the former case when the SHG sign and value are field-dependent (Figs.~\ref{fig:fig2} and~\ref{fig:fig3}) and reciprocal in the latter case when they are field-independent (Figs.~\ref{fig:fig4}b and~\ref{fig:fig4}c).

The magnetoinduced and crystallographic SHG contributions, forbidden at normal incidence, become allowed for oblique incidence when the incident/SHG wave vector $\bm k$ does not coincide with any of the crystallographic $x$, $y$, or $z$ axes. 
In fact, when the sample in the $\bm k\parallel{}y$ 
geometry is slightly rotated in positive (Fig.~\ref{fig:fig4}d) or negative (Fig.~\ref{fig:fig4}g) direction around the $x$ axis, the electric field $\bm E^{\omega}$ of the incident wave acquires nonzero projections on the two crystallographic axes ($E^{\omega}_{y}\neq{0}$, $E^{\omega}_{z}\neq{0}$) and the nonlinear polarization $P^{2\omega}_{x}$ assumes the following form 
\begin{multline}
\label{eq:eq2}
    P^{2\omega}_{x} = C^{L}_{xzzy} E^{\omega}_{z} E^{\omega}_{z} L_{y} +2 C^{B}_{xyzz}  E^{\omega}_{y} E^{\omega}_{z} B_{z} \\+ 2C_{xyz} E^{\omega}_{y} E^{\omega}_{z}.
\end{multline}
Obviously, oblique incidence enables nonreciprocal SHG due to the interference between the antiferromagnetic, magnetoinduced and crystallographic terms in Eq.~\eqref{eq:eq2}.
In the presence of interference, the absolute value of the SHG can be controlled either by inverting the applied magnetic field, $| P^{2\omega}_{x} (B_{z}) | \neq | P^{2\omega}_{x} (-B_{z}) |$, or by inverting the rotation angle, which changes the sign of $E_y$,  $| P^{2\omega}_{x} (E_{y}) | \neq | P^{2\omega}_{x} (-E_{y}) |$.

These symmetry arguments are in full agreement with the results presented in Figs.~\ref{fig:fig4}e and~\ref{fig:fig4}f for positive sample rotation, and in Figs.~\ref{fig:fig4}h and~\ref{fig:fig4}i for negative rotation. The SHG spectroscopic response changes dramatically when the sample is rotated by only the small angle of about $\pm5^{\circ}$ around the $x$ axis while keeping unchanged the incident and SHG light propagation direction 
and the magnetic field $\bm{B}\perp{}\bm{k}$.
The overall increase of the SHG intensity by about a factor of five for the rotated sample (compare Fig.~\ref{fig:fig4}b and Figs.~\ref{fig:fig4}e,h) is a clear evidence of the crystallographic SHG source becoming activated.
It is in agreement with Eq.~\eqref{eq:eq2} and manifested by the resonant and non-resonant (extended over a broad spectral range) contributions to the SHG that are independent on the magnetic field.
As a result, in the rotated samples the field-dependent and field-independent SHG contributions can interfere and the interference term varies strongly with the applied field.  
Figs.~\ref{fig:fig4}f and~\ref{fig:fig4}i show that positive and negative rotations result in opposite magnetic field dependences of the SHG.
We note that the rotational anisotropies for the both sample rotations remain similar to that for normal incidence.

The observed asymmetric shapes of the SHG spectra in the rotated samples (Figs.~\ref{fig:fig4}f and~\ref{fig:fig4}i) resemble a Fano-type resonance~\cite{fano1961effects,miroshnichenko2010fano,limonov2017fano} which, however, is inverted for positive compared to negative sample rotation.
Such Fano-shapes arise from the interference of the spectrally broad crystallographic SHG with the resonant SHG induced by the toroidal vector and magnetic field. 
The inversion of the asymmetry sign for  opposite magnetic fields is a direct evidence of the magnetic nonreciprocity.

Symmetry analysis shows that there are three main electric-dipole SHG sources, namely the crystallographic $C$ source, magnetoinduced $\bm{B}$ source, and antiferromagnetic $\bm{L}$ source which are shown schematically in Fig.~\ref{fig:fig1}c.
However, above the antiferromagnetic phase transition at $T_N=20$\,K the antiferromagnetic $\bm{L}$ source vanishes.
Nevertheless, two other $C$ and $\bm{B}$ sources in Eq.\,(2) remain allowed above the transition, and this challenged us to check whether they are capable to lead to the SHG nonreciprocity.
Such experiments were carried out and they confirmed that above $T_N$ at 25\,K there is a well-pronounced nonreciprocity due to the interference of crystalline and magnetoinduced contributions.
Results are presented and discussed in the Supplementary Section S4B. 

To conclude, we have demonstrated a new mechanism of nonlinear nonreciprocity due to the interference of second-harmonic sources induced by the applied magnetic field and the toroidal moment.
Moreover, this mechanism of nonreciprocity is supplemented by interference of the toroidal moment with the crystallographic SHG source.
In the paramagnetic phase above $T_N$, the nonlinear nonreciprocity is observed due to the crystallographic and magnetoinduced contributions to SHG.  
Our studies showed that \CBO is a very favorable platform to explore different sides of resonant  nonlinear nonreciprocity induced by a nontrivial spin order. Without any doubt such approach can be generalized to many other antiferromagnets.
Nonreciprocal effects, both linear and nonlinear, are of prime importance as efficient tools for studying electronic and magnetic structures of materials as well as for constructing technologically novel nonreciprocal optical and microwave devices.
We may add that our result will open new nonlinear degrees of freedom in the emerging field of antiferromagnetic spintronics and opto-spintronics~\cite{baltz2018antiferromagnetic,jungwirth2016antiferromagnetic,nemec2018antiferromagnetic,gomonay2014spintronics}.}

\section*{Author contributions}
J.M. performed  the experiments and analyzed the data. A.N.P. developed the theoretical model. R.M.D. derived equations and performed calculations of SHG rotational anisotropies.  D.R.Y., M.B. and R.V.P. conceived the idea for the experiment. All authors discussed the results and commented on the manuscript.  R.V.P. supervised the project.

\section*{Competing interests}
The authors declare no competing interests.

\section*{Methods}
For solving the task of registration and distinguishing different nonreciprocal contributions to the SHG processes we used a spectroscopic technique based on application of femtosecond laser pulses at 30\,kHz repetition rate.
This technique provides high sensitivity and high spectral resolution, limited only by the spectrometer for dispersing the signal.
The experimental setup is described in detail in Supplementary Section S1. 
The method was applied for the SHG study of \CBO $(xz)$-plane single-crystal samples with the incident and SHG light propagating along the $y$ axis, $\bm{k}\parallel{}y$.
The covered temperature range 1.9--25\,K includes several phase transitions between commensurate and incommensurate antiferromagnetic spin structures, as well as the antiferromagnetic--paramagnetic phase transition at $T_{N}=20$\,K~\cite{petrova2018copper}. 
The magnetic field $\bm{B}$ up to $\pm{10}$\,T was applied along the main crystallographic $x$ and $z$ axes in the Voigt geometry, $\bm{k}\perp{}\bm{B}$. Rotational anisotropies of the SHG signal were measured for crossed $\bm{E}^{\omega}\perp{}\bm{E}^{2\omega}$ and parallel $\bm{E}^{\omega}\parallel{}\bm{E}^{2\omega}$ polarizations of the incident and SHG light.
That allowed us to distinguish the symmetry-different contributions to the SHG intensity.
The anisotropies were fitted using appropriate equations, derived on the basis of the crystallographic and magnetic symmetry of \CBO as described in the Supplementary Section S3.

\section*{Acknowledgements}
We are grateful to M.A.~Gorlach and E.L.\,Ivchenko 
for useful discussions.
The samples used in our study were prepared from single crystals grown by L.N.\,Bezmaternykh.
We acknowledge the financial support by the Deutsche Forschungsgemeinschaft through the International Collaborative Research Centre 160.
A.N.P. acknowledges the partial financial support from the Russian Foundation for Basic Research Grant No.19-52-12038-NNIO\_a.
R.V.P. acknowledges the partial financial support by the RFBR Project No.19-52-12063.

\bibliography{bibliography}

\end{document}


\title{Supplementary Information:\\ Magneto-toroidal nonreciprocity of second harmonic generation}

\author{J.~Mund}
\affiliation{Experimentelle Physik 2, Technische Universit{\"a}t Dortmund, D-44221 Dortmund, Germany}
\author{D.~R.~Yakovlev}
\affiliation{Experimentelle Physik 2, Technische Universit{\"a}t Dortmund, D-44221 Dortmund, Germany}
\affiliation{Ioffe Institute, Russian Academy of Sciences, 194021 St.-Petersburg, Russia}
\author{A.~N.~Poddubny}
\affiliation{Ioffe Institute, Russian Academy of Sciences, 194021 St.-Petersburg, Russia}
\author{R.~M.~Dubrovin}
\affiliation{Ioffe Institute, Russian Academy of Sciences, 194021 St.-Petersburg, Russia}
\author{M.~Bayer}
\affiliation{Experimentelle Physik 2, Technische Universit{\"a}t Dortmund, D-44221 Dortmund, Germany}
\affiliation{Ioffe Institute, Russian Academy of Sciences, 194021 St.-Petersburg, Russia}
\author{R.~V.~Pisarev}
\email{pisarev@mail.ioffe.ru}
\affiliation{Ioffe Institute, Russian Academy of Sciences, 194021 St.-Petersburg, Russia}


\maketitle

\onecolumngrid

\subsection{\normalsize{S1. Experimental setup}}

\begin{figure*}[h]
\centering
\includegraphics[width=0.75\columnwidth]{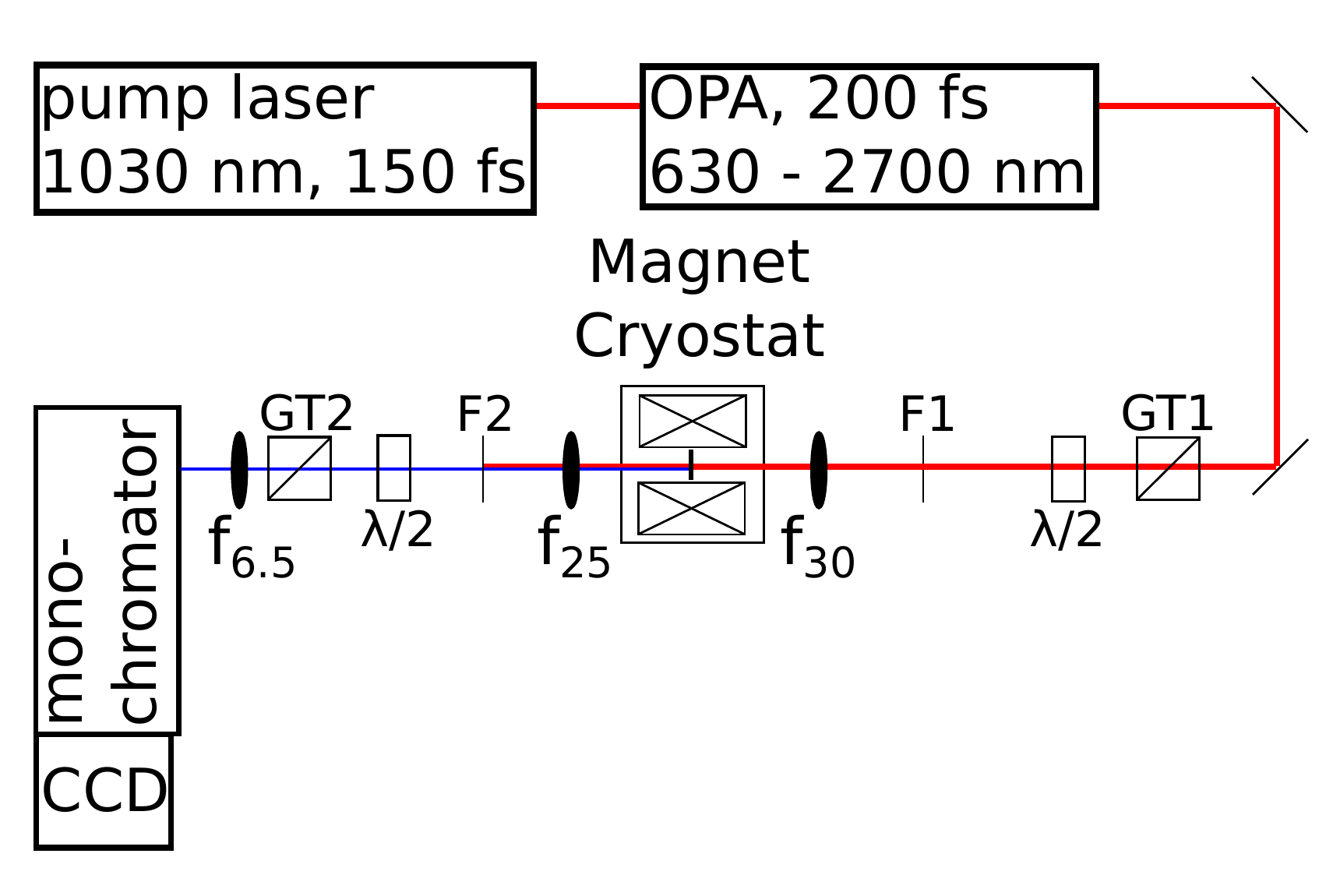}
\caption{\label{fig:setup}
\textbf{Experimental setup.}
A schematic representation of the experimental setup for SHG spectroscopy.
Optical parametric amplifier (OPA),
Glan Thompson linear polarizer (GT),
half-wave plate ($\lambda/2$),
color filter (F),
lens with $xx$-cm focal length (f$_{xx}$).   
}
\end{figure*}

The experimental setup is shown in Fig.~\ref{fig:setup}.
The exciting laser (Light Conversion) consists of an optical parametric amplifier (OPA, Orpheus) pumped by a pulsed laser (Pharos).
The OPA can be tuned in the spectral range of interest, supplying pulses of about 200\,fs duration with a full width at half maximum (FWHM) of 10\,meV.
The energy per pulse can be up to 1\,$\mu$J at a repetition rate of 30\,kHz.
The laser radiation with frequency $\omega$ and wave vector $\bm{k}^{\omega}$ hits the sample normal to its surface and is focused there into a spot of about 100\,$\mu$m diameter.

With the Glan Thompson polarizer GT1 and the half-wave plates $\lambda/2$ 
, the linear polarization of the incoming and outgoing light can be varied continuously and independently around the optical axis.
One can thus measure for any fixed polarization of $\bm{E}^{\omega}$ or $\bm{E}^{2\omega}$ the angular dependence of the other polarization.
Of special interest are the rotational anisotropies of the SHG signals for either parallel ($\bm E^{\omega} \parallel \bm E^{2\omega}$) or crossed ($\bm E^{\omega} \perp \bm E^{2\omega}$) polarizations of the laser and the SHG.
The GT2 polarizer is set to the optimal polarization of the monochromator, providing highest throughput.
The long pass filter F1 prevents SHG from optical elements to enter the monochromator and the short pass filter F2 cuts off the infrared pump light.

The detection of the SHG signal with frequency $2\omega$ and wave vector $\bm k^{2\omega} = 2 \bm k^{\omega}$ is done by a 0.5-meter monochromator (Acton, Roper Scientific) with a 1800\,grooves/mm grating in combination with a Si charge-coupled device (CCD) camera (matrix with 1340 $\times$ 400 pixels, pixel size 20\,$\mu$m) cooled by liquid nitrogen. The overall spectral resolution of the detection system at photon energies of 2.8\,eV is about 90\,$\mu$eV.

For measurements in a magnetic field, a split-coil magnet (Oxford Instruments) is used allowing field strength up to 10\,T in the Voigt configuration.

\subsection{\normalsize{S2. Symmetry analysis for the second harmonic generation}}
Here we present a symmetry analysis of the second harmonic generation in \CBO.
Our approach is based on the analysis of the magnetic symmetry of spin ordering in the $4b$ subsystem performed in Ref.~\cite{boehm2003complex}.

\subsection{A. Crystallographic symmetry}

\begin{figure*}[h!]
\centering
\includegraphics[width=1\columnwidth]{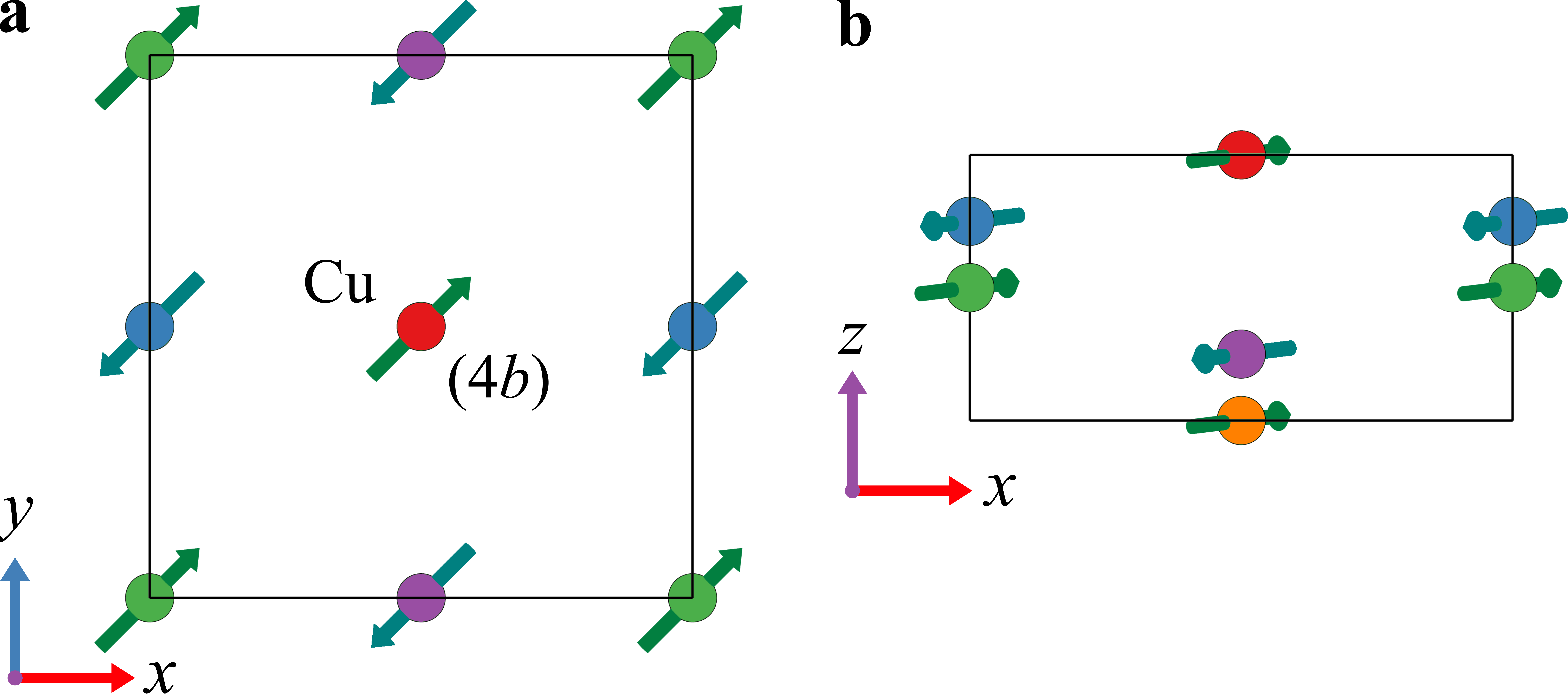}
\caption{\label{fig:Cu_CrystalStructure}
Projection of \CBO magnetic structure on $(xy)$ plane (a) and $(xz)$ plane (b).
}
\end{figure*}

Crystal symmetry of \CBO is described by the $I\bar42m$($D_{2d}^{12}$) group~\cite{martinez1971crystal,aroyo2011crystallography}.
The corresponding point group $D_{2d}$ has five classes, including identity operation $e$, $C_2$ rotation around the tetragonal axis $z\parallel c$, $S_4$ and $S_4^3$ improper rotations around $c$, two $C_2'$ rotations around the  $x\parallel a$ and $y\parallel c$ axes and two $\sigma_d$ mirror reflections in the planes $x\pm y$~\cite{dresselhaus2008group}.
The Cu$^{2+}$ ions occupy the $4b$ Wyckoff positions in the unit cell. They form two equivalent body-centered sublattices  $4b(A_1)$ and $4b(A_2$), centered at $(0,0,c/2)$ and $(a/2,0,c/4)$, as shown in Fig.~\ref{fig:Cu_CrystalStructure}.

In our experiments no SHG signals were observed in the incommensurate phase ($T>20$\,K) and we will perform analysis only in the commensurate phase, when the magnetic unit cell has the same period as the crystal unit cell.
In this case the only relevant  difference between the full nonsymmorphic spatial group $D_{2d}^{12}$ and the point group $D_{2d}$ is that the operations $\sigma_d$ and $C_{2'}$ are accompanied by the nontrivial translations by the vector $(a/2,0,3c/4)$, respectively.
Thus, these operations move Cu$^{2+}$ ions from the one $A$ sublattice to another. 
All  other symmetry operations keep 4$b$ Cu$^{2+}$ ions in the same $A$ sublattice.

The $D_{2d}$ group has five irreducible representations. Their basis functions, multiplication table and irreducible representations are presented for convenience in Table~\ref{tab:mult}.
\begin{table}[h]
\caption{Multiplication table for the irreducible representations for the $D_{2d}$ point group. Functions $x,y,z$ correspond to Cartesian projections of the polar vector; vector $\bm B$ is an axial vector (e.g. magnetic field); vector $\bm L=\bm S_{A1}-\bm S_{A2}$ describes the antiferromagnetic moment of Cu$^{2+}$ (4$b$) ions~\cite{boehm2003complex}.
}\label{tab:mult}
\begin{tabular}{|c|c|c|c|c|c|}
\hline
&$A_1$&$A_2$&$B_1$&$B_2$&$E$\\\hline 
$A_1\equiv \Gamma_1$, ${x}^2+{y}^2$ or $z^2$ &$A_1$&$A_2$&$B_1$&$B_2$&$E$\\\hline
$A_2\equiv \Gamma_2$, $B_z$ &&$A_1$&$B_2$&$B_1$&$E$; ${x}\propto B_z{y},{y}\propto - B_z{x}$\\\hline
$B_1\equiv \Gamma_3$, ${x}^2-{y}^2$ &&&$A_1$&$A_2$&$E$; ${x}\propto {x}({x}^2-{y}^2),{y}\propto {y}({y}^2-{x}^2)$\\\hline
$B_2\equiv \Gamma_4$, $2{x}{y}$ or  $z$ &&&&$A_1$&$E$; ${x}\propto {y}z,{y}\propto {x}z$\\\hline
$E\equiv \Gamma_5$, (${x},{y}$) or  (${x}z,{y}z$) or  ($B_{x},-B_{y}$) or ($L_y, L_x$) &&&&&$A_1$: $x_1x_2+y_1y_2$\\ 
 &&&&&$A_2$: $x_1y_2-y_1x_2 \equiv [\bm r_1\times\bm r_2]_z$\\
  &&&&&$B_1$: $x_1x_2-y_1y_2$ \\
    &&&&&$B_2$: $x_1y_2+x_2y_1$ \\
\hline
\end{tabular}
\end{table}

\subsection{B. Selection rules for SHG generation}
We are interested in the behavior of the incident (pump) electric field $\bm E^\omega$, second harmonic  polarization $\bm P^{2\omega} $, magnetic field $\bm B$ and the antiferromagnetic moment $\bm L=\bm S_{A1}-\bm S_{A2}\perp z$ under the crystal symmetry operations.
The components of these vectors, perpendicular to the tetragonal $c$ axis, transform according to the $E\equiv \Gamma_5$ irreducible representation .
However, specific expressions for the basis functions are different for different vectors: functions 
$E_x^\omega,P_x^{2\omega}, B_x, L_y$ transform in the same way as $x$, while the functions 
$E_y^\omega, P_y^{2\omega}, -B_y, L_x$ transform in the same way as $y$.
Hence, the symmetry properties of the antiferromagnetic moment $\bm L$ are different from both those of the polar vector $\bm E^{\omega}$ and the  axial vector $\bm B$, see also the following subsection C for more details.
The components of electric field $E^{\omega}_z$, $P^{2\omega}_z$ and magnetic field $B_z$ along the tetragonal axis are transformed according to the one-dimensional irreducible representations 
$B_2\equiv \Gamma_4$ and $A_2 \equiv \Gamma_2$, respectively.

In order to derive the selection rules for the second harmonic (SH) operations we use the multiplication Table~\ref{tab:mult} to determine the symmetry properties of the products of type $E_\alpha E_\beta$ for crystallographic SHG, $E_\alpha E_\beta L_\gamma$ for the antiferromagnetic SHG and the 
$E_\alpha E_\beta B_\gamma$ for the magnetoinduced SHG ($\alpha,\beta,\gamma=x,y,z$).
According to the Wigner-Eckart theorem~\cite{dresselhaus2008group}, we need to find the symmetrized products that transform in the same way as the SH polarization $\bm P^{2\omega}$.

We start the analysis from the crystallographic SHG. According to Table~\ref{tab:mult}, the functions 
$E_x^\omega E_z^\omega$ and $E_y^\omega E_z^\omega$ transform as the basis functions $x$ and $y$ of the irreducible representation $E$; while the product $E^\omega_xE_y^\omega$ belongs to the irreducible representation $B_2$. Hence, these two types of products contribute to the corresponding components of SHG polarization $P_x,P_y$ and $P_z$ with different weights 
$C_{xyz}$ = $C_{yxz}$ and $C_{zyx}$ = $C_{zxy}$  :
\begin{align}
P^{2\omega}_x&=2C_{xyz}E^\omega_yE_z^\omega\:,\label{eq:cryst}\\
P^{2\omega}_y&=2C_{xyz}E^\omega_xE_z^\omega\:,\nonumber\\
P^{2\omega}_z&=2C_{zyx}E^\omega_xE_y^\omega\nonumber\:.
\end{align}

\begin{table}[h!]
\caption{Multiplication table to derive the selection rules for magnetoinduced and antiferromagnetic SHG. Left column indicate the symmetry properties of the products of electric field components $E_\alpha^\omega E_\beta^\omega$.
Top row shows the transformation rules for the projections of magnetic field and antiferromagnetic moment $B_\gamma$ and $L_\gamma$.
The cells indicate the transformation properties of the products $E_\alpha E_\beta B_\gamma$ and $E_\alpha E_\beta L_\gamma$.
Blue color indicates contributions to Eqs. (1) and (2) of the main text. 
}\label{table:all}
{\small\begin{tabular}{|c|c|c|}
\hline
&$A_2: B_z$&$E: (B_x,-B_y)\text{ or } (L_y, L_x)$\\\hline 
$A_1: E_x^{\omega,2}+E_y^{\omega,2}$ &$A_2$&{$E: P^{2\omega}_x= (E_x^{\omega,2}+E_y^{\omega,2})(C_{xxxx}^BB_x+C_{xxxy}^LL_y);$} \\
&&{$P^{2\omega}_y=(E_x^{\omega,2}+E_y^{\omega,2})(-C_{xxxx}^BB_y+C_{xxxy}^LL_x)$}\\\hline
$A_1: E_z^{\omega,2}$ &$A_2$&{\color{blue}$ P^{2\omega}_x= E_z^{\omega,2}(C_{xzzx}^BB_x+C_{xzzy}^LL_y);$}\\
&&{$P^{2\omega}_y= E_z^{\omega,2}(-C_{xzzx}^BB_y+C_{xzzy}^LL_x)$}\\\hline
$B_1: E_x^{\omega,2}-E_y^{\omega,2}$ &{$P^{2\omega}_z=C^B_{zxxz}(E_x^{\omega,2}-E_y^{\omega,2}) B_z$}&
{$P^{2\omega}_x= (E_x^{\omega,2}-E_y^{\omega,2})(-C_{yyyx}^BB_y+C_{yyyy}^LL_x);$}\\
&&{$ P^{2\omega}_y= -(E_x^{\omega,2}-E_y^{\omega,2})(C_{yyyx}^BB_x+C_{yyyy}^LL_y)$}\\\hline
$B_2: E_x^\omega E_y^\omega $ &$B_1$& {$P^{2\omega}_x= 2E_x^{\omega}E_y^{\omega}(-C^B_{yxyx}B_y+C^L_{yxyy}L_x);$}\\
 && {$ P^{2\omega}_y=2E_x^{\omega}E_y^{\omega}(C^B_{yxyx}B_x+C^L_{yxyy}L_y)$}\\
\hline
$E: (E_x^\omega E_z^\omega ,E_y^\omega E_z^\omega )$ &{\color{blue}${ P^{2\omega}_x=2C_{xyzz}^BE_y^{\omega}E_z^{\omega}B_z}$}& {$ P^{2\omega}_z= 2E_x^\omega E_z^\omega (-C_{zyzx}^BB_y+C_{zyzy}^LL_x)$
}\\&$P^{2\omega}_y= -2C_{xyzz}^BE_x^{\omega}E_z^{\omega}B_z$&{$+
2E_y^\omega E_z^\omega (C_{zyzx}^BB_x+C_{zyzy}^LL_y)$}\\
\hline
\end{tabular}}
\end{table}

Now we proceed to the antiferromagnetic and magnetoinduced SHG. To this end, we first write the transformation products of all the possible products $E^\omega_\alpha E^\omega_\beta$ of electric field projections in the left column of Table~\ref{table:all}. 
Top row of the table lists the projections of magnetic field $\bm B$ and antiferromagnetic vector $\bm L$.
Next, we fill the cells of the multiplication Table~\ref{table:all} by the symmetry properties of different possible products $E_\alpha E_\beta B_\gamma$ and $E_\alpha E_\beta L_\gamma$ of these projections.  
The combinations  transforming in the same way as $\bm P^{2\omega}$ yield the desired selection rules for the SHG. 
The combinations highlighted by the blue color yield Eq.\,(2) 
for the SHG in the main text.
We note, that the contributions to the SHG linear in the ferromagnetic moment $\bm M=\bm S_1+\bm S_2$ can also be obtained from the Table~\ref{table:all}. The reason is that vectors  $\bm B$ and $\bm M$ transform in the same way, so the symmetry of the corresponding tensor $C^M_{ijkl}$ is the same as for $C^B_{ijkl}$.

\subsection{C. Antiferromagnetic contribution to the SHG\\ and its relation to the toroidal moment}
In this section we analyze in more detail the contribution to the SHG, linear in antiferromagnetic moment $\bm L$.
It is described by the tensor components $C_{xzzy}^L=C_{yzzx}^L$, or, explicitly,
\begin{equation}\label{eq:PL}
P^{2\omega}_x= C_{xzzy}^LE_z^{\omega,2}L_y,\quad P^{2\omega}_y= C_{xzzy}^LE_z^{\omega,2}L_x\:.
\end{equation} 
Since the intensity of light along the tetragonal axis $E_z^{\omega,2}$ is an invariant in the group $D_{2d}^{12}$, Eq.~\eqref{eq:PL} follows directly from the basis functions of the representation $\Gamma_5\equiv E$, i.e. $P^{2\omega}_x$ transforms in the same way as $L_y$ and 
$P^{2\omega}_y$ transforms in the same way as $L_x$. 
Our goal is to get better understanding of  this linear relation between the SHG and the antiferromagnetic vector.
It is instructive to link  the vector
\begin{equation} \label{eq:T0}
\bm T=L_y\hat{\bm x}+L_x\hat{\bm y}
\end{equation} to  the toroidal  vector~\cite{spaldin2008toroidal}\:
\begin{equation}\label{eq:t}
\bm T=\frac{1}{2}\sum\limits_{j}\bm r_j\times \bm S_j\:.
\end{equation} 
The toroidal moment Eq.~\eqref{eq:t} behaves as a vector, and hence the effects
proportional to $\bm P^{2\omega}\propto \bm t$ should be allowed by the point group symmetry.
However, it is not fully clear at the moment \cite{spaldin2008toroidal}, how the expression Eq.~\eqref{eq:t}, written for a localized spin distribution, can be generalized  to a periodic crystal.
It has been suggested~\cite{tokura2018nonreciprocal} that the observation of SHG in multiferroic FeGaO$_3$ is related to the toroidal moment 
\begin{equation}\label{eq:t1}
\bm E^{2\omega}\propto \bm T=\bm P\times \bm M\:, \quad T_x=P_zM_y,\quad T_y=-P_zM_x,
\end{equation}
where $\bm M$ is the ferrimagnetic moment (resulting from unbalance between two antiferromagnetically-coupled sublattices) and $\bm P\parallel z$ is the electric polarization.
However, the symmetry structure of our Eq.~\eqref{eq:PL} is different from Eq.~\eqref{eq:t1}. The reason is the analysis of toroidal moment in 
\cite{tokura2018nonreciprocal}  has been performed for the different crystal structure. Therefore, it is not clear why results for toroidal moment from \cite{tokura2018nonreciprocal,arima2019} should apply to \CBO.

In our case the tensor linking $\bm P^{2\omega}$ and $\bm L$ is symmetric, while in Eq.~\eqref{eq:t1} this tensor is antisymmetric.
We explain this difference by the nonsymmorphic nature of the symmetry group $D_{2d}^{12}$ of CuB$_{2}$O$_{4}$. 
For a better understanding of Eq.~\eqref{eq:PL} we write the spin density in the antiferromagnetic phase as
\begin{equation}
\bm l(\bm r)=P_z(\bm r)\bm S,\quad P_z(\bm r)=\sum\limits_{j}[\delta(\bm r-\bm r_{A_1,j})-\delta(\bm r-\bm r_{A_2,j})]\:, S_z=0.
\end{equation} 
where the summation runs over the $4b$ Cu$^{2+}$ ions in sublattices $A_1$ and $A_2$. 
The symmetry properties of $\bm l(\bm r)$ are determined by the product of the representations governing the transformation of $P_z(\bm r)$ and the spin $\bm S$.
The spin components $S_x$ and $S_y$  behave under the point symmetry operations as the $x$ and $-y$, respectively, i.e. in the same way as a pseudovector in a $D_{2d}$ point group~\cite{dresselhaus2008group}. 
The function $P_z(\bm r)$ belongs to the irreducible representation $\Gamma_2\equiv A_2$, i.e. behaves under the symmetry operations as a $z$-component of a pseudovector. 
This follows from the fact that two $A$ sublattices are linked by the rotations $C_2'$ and the mirror reflections $\sigma_d$, that require nontrivial translation in the $D_{2d}^{12}$ spatial group.
Thus, the $\sigma_d$ operation swaps the sublattices and changes the sign of $P_z$, in the same way changes the sign of a pseudovector component $B_z$.
According to Table \ref{tab:mult}, the product of the $A_2$ representations with the basis function $P_z$ and $E$ with the basis functions $X,Y$ is the $E$ representation with the basis functions ${x}\propto -P_zY,\quad {y}\propto  P_zX$.
Substituting $Y=-S_y$ and $X=S_x$, we find that $P_zS_y\equiv L_y$ transforms as $x$ and $PS_x\equiv L_x$ transforms as $y$.
This is in full agreement with Table \ref{tab:mult} and Eq.~\eqref{eq:PL}.
To summarize, we find that Eq.~\eqref{eq:t1} in our system can be formally generalized to
\begin{equation}\label{eq:t2}
    \bm E^{2\omega}\propto \bm T=L_y\hat{\bm x}+L_x\hat{\bm y}=P_z\otimes \bm S,
\end{equation}
where the vector product is replaced by the direct one.
This equation can be seen as a generalization of the toroidal moment Eq.~\eqref{eq:t1} for the nonsymmorphic $D_{2d}^{12}$ crystal structure of CuB$_{2}$O$_{4}$.

\subsection{D. Toroidal and quadrupolar contributions to the linear magnetoelectric effect}
It is instructive to analyze the  static  magnetoelectric effect~\cite{dzyaloshinskii1959magnetoelectrical}, where the polarization at zero frequency $\bm P$ is proportional to the antiferromagnetic momentum $\bm L$ and the applied magnetic field $\bm B$. 
Using the transformation rules for the vectors $\bm P$, $\bm L$, $\bm B$, discussed above and summarized in the Table~\ref{tab:mult}, we find the magnetoelectric relationship   in \CBO in the following form
\begin{equation} 
P_j=\alpha^{\rm (em)}_{jk}B_k\:.\label{eq:PB2}
\end{equation} 
where the magnetoelectric tensor $\alpha^{\rm (em)}$ reads
\begin{equation}
\alpha^{(em)}=\begin{pmatrix} 0&0&u L_x\\0&0&-u L_y\\ v L_x&-v L_y&0\end{pmatrix}
\end{equation}
and is described by two independent coefficients $u$ and $v$. Following~\cite{spaldin2008toroidal}, we separate the magnetoelectric tensor into symmetric and antisymmetric parts and rewrite  Eq.~\eqref{eq:PB2}  as 
\begin{equation} 
\bm P=-\bm t\times \bm B+Q\bm B\:.
\end{equation} 
Here the vector 
\begin{equation} \label{eq:p3}
\bm t=\frac{u-v}{2}(\bm e_x L_y+\bm e_y L_x)\:,
\end{equation} 
dual to the antisymmetric part of the tensor $\alpha^{(em)}$, is the toroidal moment vector. It is proportional to the toroidal moment vector $\bm T$ Eq.~\eqref{eq:T0}, derived  in the following subsection in a different fashion.
The symmetric part of the antiferromagnetic  tensor
\begin{equation} 
Q_{jk}=\frac{1}{2}(\alpha^{(em)}_{jk}+\alpha^{(em)}_{kj})=\frac{u+v}{2}
 \begin{pmatrix}
 0&0&L_x\\0&0&-L_y\\L_x&-L_y&0
\end{pmatrix}
\end{equation} 
describes the magnetic quadrupole response.
Comparing Eq.~\eqref{eq:PL} with Eq.~\eqref{eq:p3} we again confirm that 
the observed SHG contribution due to antiferromagnetic moment can be written as
\begin{equation}
\bm P^{2\omega}\propto\bm t (E_z^\omega)^2\:,
\end{equation}
i.e. it is proportional to the toroidal vector.

\subsection{\normalsize{S3. Analytical expressions for rotational anisotropies of SHG}}

The electric field of incident light $\bm{E}^{\omega}$ with polarization in $xz$ plane ($E^{\omega}_{x}, E^{\omega}_{z} \ne 0$, $E^{\omega}_{y} = 0$) is given by
\begin{equation}
\label{eq:E_components}
\begin{cases}
E^{\omega}_{x} = E^{\omega} \cos{\phi},
\\
E^{\omega}_{z} = E^{\omega} \sin{\phi},
\end{cases}
\end{equation}
in which $x$, $z$ indicate laboratory coordinates, and $\phi$ represents the azimuthal rotational angle.
$E^{\omega}_{x}$ and $E^{\omega}_{z}$ are Cartesian components of the electric field of incident light.
$B_{x}, B_{z} \ne 0$, $B_{y}=0$.

The SHG polarization components $P^{2\omega}_{x}$ and $P^{2\omega}_{z}$ in this case in \CBO according to Table~\ref{table:all} have the form
\begin{equation}
\begin{cases}
P^{2\omega}_{x} = ( C^{B}_{xxxx} B_{x} + C^{M}_{xxxx} M_{x} + C^{L}_{xxxy} L_{y} - C^{M}_{yyyx} M_{y} + C^{L}_{yyyy} L_{x} ) E^{\omega}_{x} E^{\omega}_{x}\\
\hspace{16em} + ( C^{B}_{xzzx} B_{x} + C^{M}_{xzzx} M_{x} + C^{L}_{xzzy} L_{y} ) E^{\omega}_{z} E^{\omega}_{z},
\\
P^{2\omega}_{z} = 2 (C^{L}_{zyzy} L_{x} - C^{M}_{zyzx} M_{y}) E^{\omega}_{x} E^{\omega}_{z} + ( C^{B}_{zxxz} B_{z} + C^{M}_{zxxz} M_{z} ) E^{\omega}_{x} E^{\omega}_{x},
\end{cases}    
\end{equation}
where the ferromagnetic component $C^{M}_{ijkl} E^{\omega}_{j} E^{\omega}_{k} M_{l}$ is taken into account as described in Supplementary Section S2B.

Angular dependencies of SHG polarization can be obtained using the Eq.~\eqref{eq:E_components} in the form
\begin{equation}
\label{eq:P_components}
\begin{cases}
P^{2\omega}_{x}(\phi) = \left[ ( C^{B}_{xxxx} B_{x} + C^{M}_{xxxx} M_{x} + C^{L}_{xxxy} L_{y} - C^{M}_{yyyx} M_{y} + C^{L}_{yyyy} L_{x} ) \cos{^2\phi}\right. \\
\hspace{14.6em} \left. + ( C^{B}_{xzzx} B_{x} + C^{M}_{xzzx} M_{x} + C^{L}_{xzzy} L_{y} ) \sin{^2\phi} \right] E^{\omega} E^{\omega},
\\
P^{2\omega}_{z}(\phi) = \left[ 2 (C^{L}_{zyzy} L_{x} - C^{M}_{zyzx} M_{y}) \cos{\phi} \sin{\phi} + ( C^{B}_{zxxz} B_{z} + C^{M}_{zxxz} M_{z} ) \cos{^2\phi} \right] E^{\omega} E^{\omega}.
\end{cases}    
\end{equation}

The parallel (perpendicular) SHG polarization can be defined as
\begin{equation}
\begin{cases}
P^{2\omega}_{\parallel}(\phi) = P^{2\omega}_{x}(\phi) \cos{\phi} + P^{2\omega}_{z}(\phi) \sin{\phi},
\\
P^{2\omega}_{\bot}(\phi) = - P^{2\omega}_{x}(\phi) \sin{\phi} + P^{2\omega}_{z}(\phi) \cos{\phi},
\end{cases}    
\end{equation}
in which $\parallel$ ($\bot$) indicates the parallel (perpendicular) component.

The intensity of parallel (perpendicular) SHG can be obtain as
\begin{equation}
\begin{cases}
I^{2\omega}_{\parallel}(\phi) = \left[ P^{2\omega}_{\parallel}(\phi) \right]^2,
\\
I^{2\omega}_{\bot}(\phi) = \left[ P^{2\omega}_{\bot}(\phi) \right]^2.
\end{cases}
\end{equation}

\subsection{\normalsize{S4. Experimental results}}

\subsection{A. SHG spectra for $\pm \bm{B}$ from 0 to 10\,T at $T$=6\,K}

\begin{figure*}[h!]
\centering
\includegraphics[width=1\columnwidth]{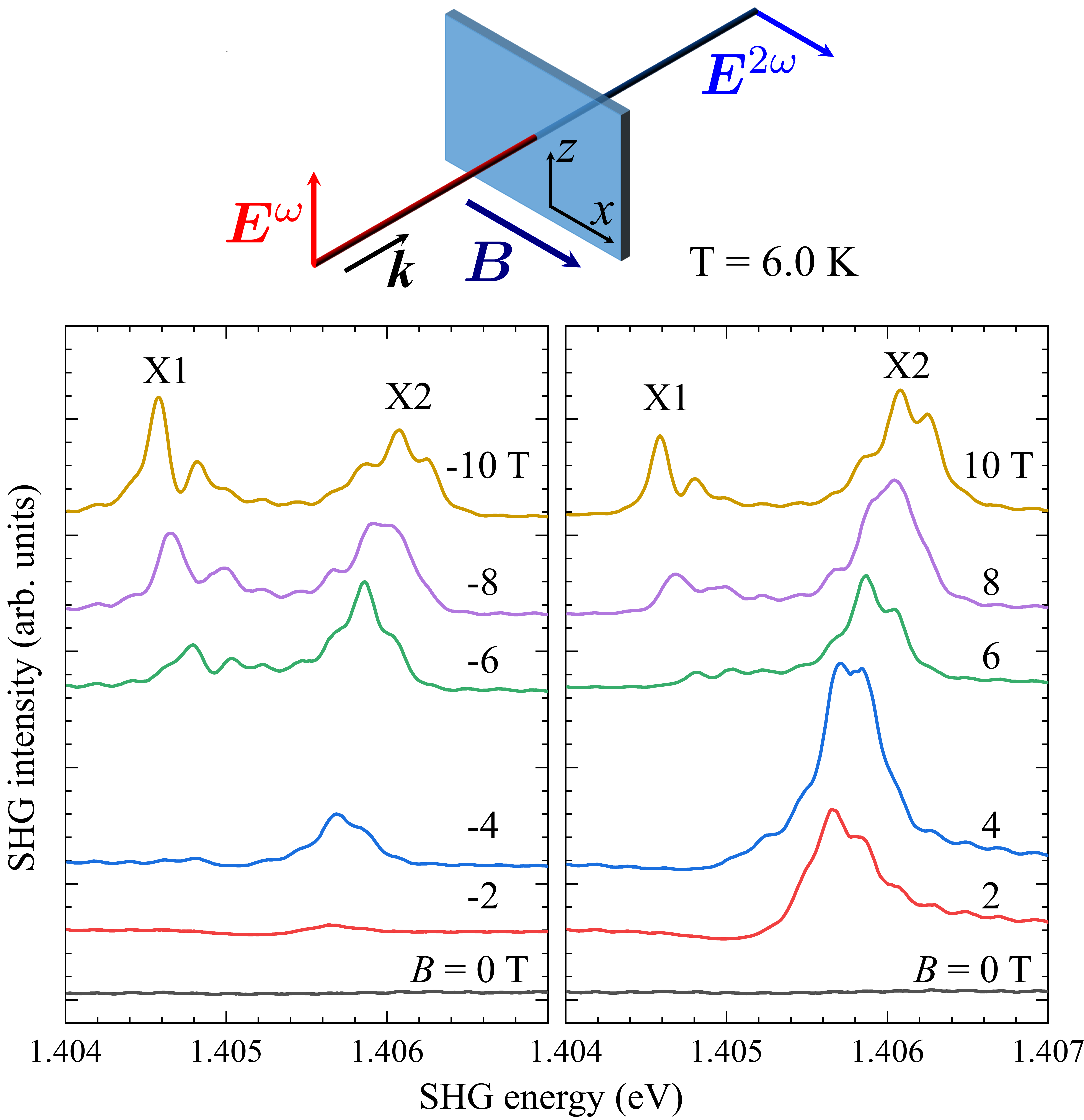}
\caption{\label{fig:S2}
SHG spectra for $\pm \bm{B}$ from 0 to 10\,T at $T$=6\,K.
}
\end{figure*}

Figure~\ref{fig:S2} shows evolution of the SHG spectra in \CBO at $T$=6\,K for the magnetic field $\pm \bm{B}$ from 0 to 10\,T applied along the $x$ axis.
No SHG is observed at zero field when the $4b$ spin system is in the incommensurate phase in accordance with the phase diagram in Fig.\,3a. 
These spectra confirm complementarity and reliability of the results at $T$=4\,K shown in Fig.\,3b.
Field increase leads to evolution of the both group of lines marked as $X1$ and $X2$ but in different way for the positive and negative field.
In the positive field $X2$ group dominates in the SHG spectra, whereas in the negative field the SHG intensity of $X1$ group overcomes that of the $X2$ group.
Such complex behaviour of SHG lines is definitely related to their microscopic nature but such an analysis is outside the scope of this article.    

\subsection{B. Nonreciprocity of the SHG spectra above $T_N$=20\,K}

\begin{figure*}[h]
\centering
\includegraphics[width=1\columnwidth]{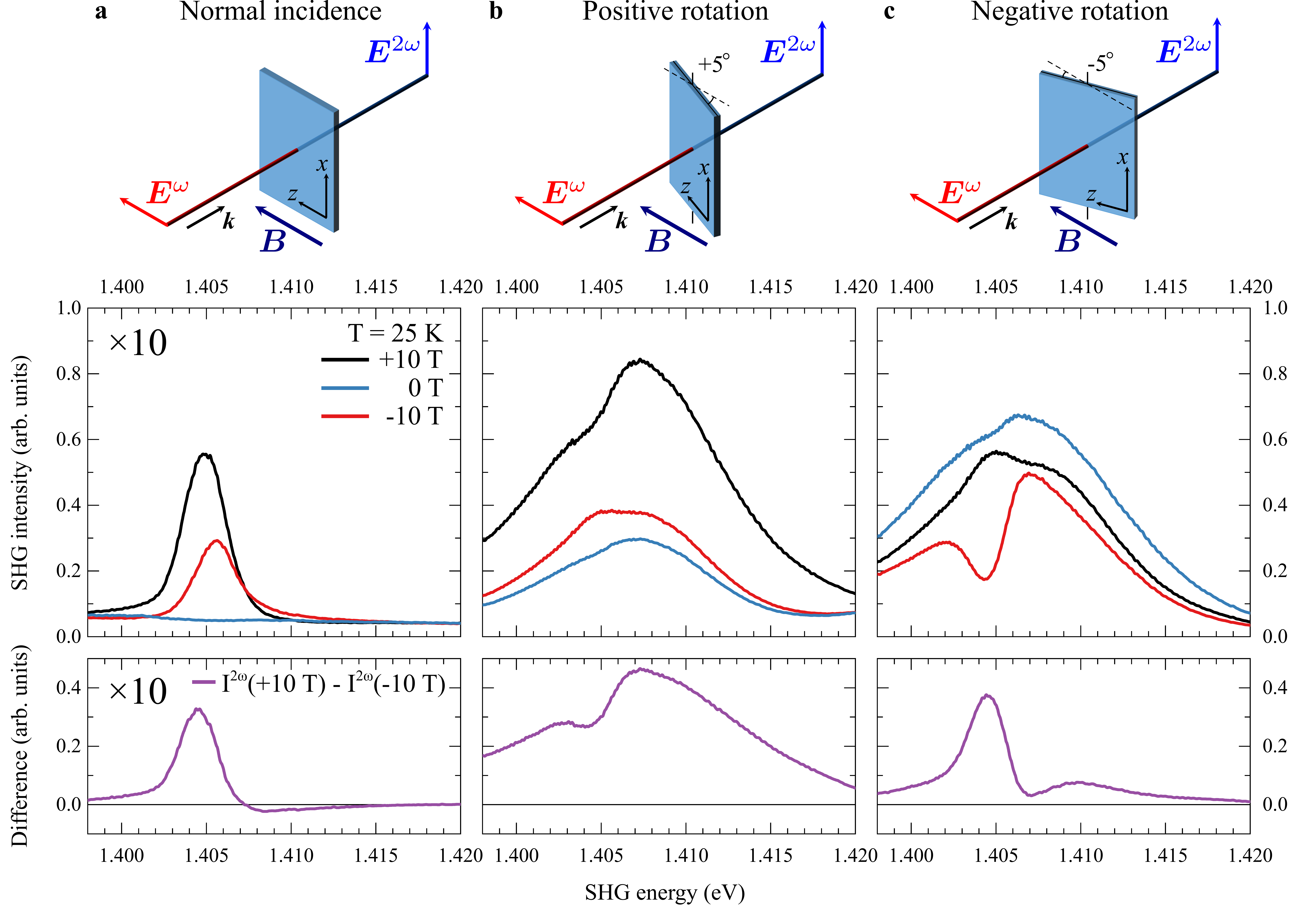}
\caption{\label{fig:S3}
SHG spectra at $T$=25\,K.
}
\end{figure*}
 
Symmetry analysis shows that in the electric-dipole approximation there are three main SHG sources, namely the $C$ source related the the noncentrosymmetric crystal structure, the magnetic-field induced $B$ source, and the antiferromagnetic $L$ source.
All of them are shown schematically in Fig.\,1c and in Eq.~(2).
Their transformation properties are different as shown in Tables~\ref{tab:mult} and~\ref{table:all} presented above in Supplementary Information. 
It was a challenging task to verify whether the $C$ and $B$ sources survive in the paramagnetic phase above $T_N$=20\,K when the antiferromagnetic source $L$ vanishes. 

To solve this challenging task experiments were carried at $T$=25\,K and results are shown in Fig.~\ref{fig:S3}. 
Experiments were carried out at normal incidence and with the sample rotated around the $x$ axis in positive and negative direction similar to results presented in Fig.\,2. 
It should be noted first of all that a dramatic drop of the SHG intensity above $T_N$ occurs. 
In order to register the signal we had to increase by about an order of magnitude the intensity of the incident beam and to perform measurements in the largest available field $\pm{B}$=10\,T. 
This allowed us to obtain reliable and convincing results.
Fig.~\ref{fig:S3}a shows that at normal incidence when crystallographic contribution is symmetry forbidden there is no SHG signal in zero field.
However, reliable signals are observed in magnetic field, which are different for opposite fields demonstrating another side of magnetoinduced SHG nonreciprocity.
SHG intensity is strongly increased for the tilted sample due to emergence of crystallographic off-resonant and resonant SHG contributions. 
The interference between the crystallographic and field-induced SHG is well pronounced in Figs.~\ref{fig:S3}b and~\ref{fig:S3}c, and in particular for the negative-tilted sample.

\bibliography{supbib}